\documentclass[11pt]{article}
\usepackage[english]{babel}
\usepackage{amsmath,amssymb,mathtools,amsthm}
\usepackage{cite}
\usepackage{xcolor}

\newcommand{\1}{{\rm 1\hspace*{-0.4ex}%
\rule{0.1ex}{1.52ex}\hspace*{0.2ex}}}
 \expandafter\let\csname
equation*\endcsname\relax \expandafter\let\csname endequation*\endcsname\relax

\usepackage[english]{babel}
\usepackage{amsmath,amssymb,mathtools,amsthm}

\allowdisplaybreaks

\newcommand{\SE}{Schr\"{o}dinger equation}

\newcommand{\half}{\frac{1}{2}}
\newcommand{\ud}{\mathrm{d}}

\newcommand{\R}{\mathbb{R}}
\renewcommand{\vec}{\mathbf}
\renewcommand{\epsilon}{\varepsilon}
\renewcommand{\imath}{\mathrm{i}}

\def\dddot#1{\mathinner{\buildrel\vbox{\kern5pt\hbox{...}}\over{#1}}}

\newcommand{\di}{\displaystyle}
\newcounter{rmk}
\renewcommand{\thermk}{\arabic{rmk}}
\newenvironment{remark}%
{\refstepcounter{rmk}\vspace{6pt}\noindent\ignorespaces\textbf{Remark
\thermk:}}{\vspace{6pt}\par}

\begin{document}

\title{Quantization of quadratic Li\'enard-type equations by preserving Noether symmetries}
\author{G. Gubbiotti$^1$ and M.C. Nucci$^{2}$}
\date{$^1$ Dipartimento di Matematica e Fisica,
Universit\`{a} degli Studi Roma Tre,\\ \& INFN Sezione di Roma Tre, 00146 Roma,
Italy\\[0.2cm]
$^2$ Dipartimento di Matematica
 e Informatica, Universit\`a degli Studi di Perugia, \\\& INFN Sezione di Perugia, 06123 Perugia,
 Italy}
%\ead{}
\maketitle

\begin{abstract}
The classical quantization of a family of a quadratic Li\'{e}nard-type equation
(Li\'{e}nard II equation) is achieved by a quantization scheme (M.~C. Nucci.
{\em Theor. Math. Phys.}, 168:994--1001, 2011) that preserves the Noether point
symmetries of the underlying Lagrangian in order to construct the Schr\"odinger
equation. This method straightforwardly yields the Schr\"odinger equation as
given in (A.~Ghose~Choudhury and Partha Guha. {\em J. Phys. A: Math. Theor.},
46:165202, 2013).
\end{abstract}

\section{Introduction}

In \cite{c-iso} it was inferred that Lie symmetries should be preserved if a
consistent quantization is desired. In \cite{Goldstein80} [ex. 18, p. 433] an
alternative Hamiltonian for the simple harmonic oscillator was presented. It is
obtained by applying a nonlinear canonical transformation to the classical
Hamiltonian of the harmonic oscillator. That alternative Hamiltonian was used
in \cite{Nucci2013} to demonstrate what nonsense the usual quantization
schemes\footnote{Such as normal-ordering \cite{Bjorken1964,Louisell1990} and
Weyl quantization \cite{Weyl1927}.} produce. In \cite{gallipoli10} a
quantization scheme that preserves the Noether symmetries was proposed and
applied  to Goldstein's example in order to derive the correct Schr\"odinger
equation. In \cite{Bregenz11}  the same quantization scheme was applied in
order to quantize the second-order Riccati equation while in \cite{Nucci2012}
the quantization of Calogero's goldfish system was achieved.
 In \cite{GN_liensch} it was shown that this
method straightforwardly yields the Schr\"odinger equation in the momentum
space of  a Li\'{e}nard-type nonlinear oscillator as given in
\cite{Chandrasekar2012}.\\

 If a system of second-order equations is considered, i.e.
\begin{equation}
\vec{\ddot{x}}(t)= \vec{F}(t,\vec{x},\vec{\dot{x}}), \quad \vec{{x}}\in\R^N,
\label{systsec}
\end{equation}
that comes from a variational principle with a Lagrangian of first order, then
the method that was first proposed in \cite{gallipoli10} consists of the
following steps:
\begin{description}
\item[Step I.] Find the Lie symmetries of the Lagrange equations
$$\Upsilon=W(t,\vec{x})\partial_t+\sum_{k=1}^{N}W_k(t,\vec{x})\partial_{x_k}$$
\item[Step II.]  Among them find the Noether  symmetries
$$\Gamma=V(t,\vec{x})\partial_t+\sum_{k=1}^{N}V_k(t,\vec{x})\partial_{x_k}$$
This may require searching for the Lagrangian yielding the maximum possible
number of Noether  symmetries \cite{laggal,CP07Rao1JMP,nuctam_1lag}.
\item[Step III.]  Construct the Schr\"odinger equation\footnote{We assume
 $\hbar=1$ without loss of generality.} admitting these Noether
symmetries as Lie symmetries, namely  \begin{equation}2i\psi_t+\sum_{k,j=1}^{N}
f_{kj}(\vec{x})\psi_{x_jx_k}+
\sum_{k=1}^{N}h_k(\vec{x})\psi_{x_k}+f_0(\vec{x})\psi=0
\label{sch}\end{equation} admitting the Lie symmetries
$$\Omega=V(t,\vec{x})\partial_t+\sum_{k=1}^{N}V_k(t,\vec{x})\partial_{x_k}
+G(t,\vec{x},\psi)\partial_{\psi}$$ without adding any other symmetries apart
from the two symmetries that are present in any linear homogeneous partial
differential equation\footnote{In the following we will refer to those two
symmetries as the homogeneity and linearity symmetries.}, namely
$${\psi}\partial_{\psi}, \quad \quad \alpha(t,\vec{x})\partial_{\psi},$$
where $\alpha=\alpha(t,\vec{x})$ is any solution of the Schr\"odinger equation
\eqref{sch}.
\end{description}
\strut\hfill\\ If the system \eqref{systsec}  is linearizable by a point
transformation, and it possesses the maximal number of admissible Lie point
symmetries, namely $N^2+4N+3$, then in \cite{Gonzalez1983,Gonzalez1988} it was
proven that the maximal-dimension Lie symmetry algebra of a system of $N$
equations of second order is isomorphic to $sl(N+2,\R)$, and that the
corresponding Noether symmetries generate a $(N^2+3N+6)/2$-dimensional Lie
algebra $g^V$ whose structure (Levi-Mal\'cev decomposition and realization by
means of a matrix algebra) was determined. It was also proven that the
corresponding linear system is
\begin{equation}
\vec{y}''(s)+2A_1(s)\cdot\vec{y}'(s)+ A_0(s)\cdot\vec{y}(s)+\vec{b}(s)=0,
\label{linsys}
\end{equation}
 with the condition
\begin{equation}
A_0(s)=A_1'(s)+A_1(s)^2+a(s)\1\,,
\end{equation}
where $A_0,A_1$ are $N\times N$ matrices, and $a$ is a scalar function.\\
\\Consequently if  system \eqref{systsec} admits $sl(N+2,\R)$ as Lie symmetry algebra
then in \cite{GN_liensch} we reformulated the algorithm that yields the
Schr\"odinger equation as follows:
\begin{description}
\item[Step 1.] Find the linearizing transformation which
does not change the time, as prescribed in non-relativistic quantum mechanics.
\item[Step 2.]
Derive the Lagrangian  by applying the linearizing transformation to the
standard Lagrangian of the corresponding linear system \eqref{linsys}, namely
that that admits the maximum number of Noether symmetries\footnote{In
\cite{Gonzalez1988}  it was shown that any diffeomorphism between two systems
of second-order differential equations takes Noether symmetries into Noether
symmetries, and therefore the Lagrangian is unique up to a diffeomorphism.}.
\item[Step 3.] Apply the linearizing transformation
 to the Schr\"{o}dinger equation  of the
corresponding classical linear problem. This yields the Schr\"odinger~ equation
corresponding to system \eqref{systsec}.
\end{description}
This quantization is consistent with the classical properties of the system,
namely  the Lie symmetries of the obtained Schr\"{o}dinger equation correspond
to the Noether symmetries admitted by the Lagrangian of system
\eqref{systsec}.\\
\strut\hfill\\
In \cite{Partha2013}  the quantization of the following quadratic
Li\'enard-type equation (also called Li\'enard II equation):
\begin{equation}
\ddot{x} + f(x) \dot{x}^2 + g(x) = 0, \label{lienii}
\end{equation}
where  $f(x), g(x)$ are arbitrary smooth functions of $x$, was tackled, and the
method of the Jacobi Last Multiplier \cite{Jacobi45,Whittaker} was applied in
order to find a Lagrangian without acknowledging that it was already known
\cite{nuctam_1lag} as detailed in the following Remark.

\begin{remark}  The Li\'{e}nard II equation \eqref{lienii} is a special
 case\footnote{With the substitutions $\partial\phi/\partial t=0$,
  $\partial\phi/\partial x=2 f(x)$ and
$B(t,x)=g(x)$.} of the following class of second-order ordinary differential
equations\footnote{With $\phi(t,x)$ and $B(t,x)$ arbitrary functions of $t$ and
$x$.}:
\begin{equation}
\ddot{x} + \half \frac{\partial\phi(t,x)}{\partial x}\dot{x}^2
+\frac{\partial\phi(t,x)}{\partial t} \dot{x} + B(t,x)=0. \label{eulerjacobi}
\end{equation}
that was studied by Euler \cite{Euler1769} [Sect. I,  Ch. VI, \S\S 915 ff.]. In
\cite{Jacobi45}, Jacobi found that the last multiplier is:
\begin{equation}
M=e^{\phi(t,x)}. \label{jlmej}
\end{equation}
In \cite{nuctam_1lag} the corresponding Lagrangian was found to be:
\begin{equation}
L_{\text{EJ}} = \frac{e^{\phi(t,x)}}{2} \dot{x}^2+ \int^{x}
e^{\phi(t,\xi)}B(t,\xi)\ud \xi. \label{lagrej}
\end{equation}
Consequently a Lagrangian for the Li\'enard II equation \eqref{lienii} is given
by:
\begin{equation}
L_{\text{L-II}} = \frac{\dot{x}^2}{2}\, e^{2 \int^{x}f(\xi)\ud \xi}+ \int^{x}
g(\xi)e^{2 \int^{\xi}f(\eta)\ud \eta}\ud \xi. \label{lagrlienii}
\end{equation}

\end{remark}
In \cite{LakshmananLienII2013} the Lie point symmetry classification of
Li\'{e}nard II equation \eqref{lienii} was performed. In particular it was
shown that the following family of equations
\begin{equation}
\ddot x+ \di\frac{h''}{h'}\dot x^2 +{\lambda}\di \frac{h}{h'}+\di\frac{A}{h'
h^3}=0, \label{lieniilingen}
\end{equation}
where $\lambda, A\in \R$ are arbitrary constants and $h(x)$ is an arbitrary
smooth function of $x$, admits a three-dimensional Lie symmetry algebra
$\mathrm{sl}(2,\R)$ if $A\neq 0$, and an eight-dimensional Lie symmetry algebra
$\mathrm{sl}(3,\R)$ if $A=0$. In \cite{LakshmananLienII2013} it was also proven
that if $\lambda=\omega^2>0$ then \eqref{lieniilingen} becomes:
\begin{equation}
\ddot{x} + \frac{h''}{h'}  \dot{x}^2 + \omega^2 \frac{h}{h'} +\di\frac{A}{h'
h^3}=0. \label{lieniilinom}
\end{equation}
that is
 isochronous\footnote{We do not consider equation
\eqref{lieniilingen} with $\lambda\leq 0$, since the isochronous property of
\eqref{lieniilinom} would have been lost.} since the point transformation:
\begin{equation}
\xi=h(x) \label{lintoho}
\end{equation} transforms \eqref{lieniilinom} into the isotonic
oscillator
\begin{equation}
\ddot \xi + \omega^2 \xi+ \frac{A}{\xi^3} =0.
\end{equation}
If $A=0$ then equation \eqref{lieniilinom} is transformed by means of
\eqref{lintoho} into an harmonic oscillator with frequency $\omega$.\\
Equation \eqref{lieniilinom} with $A=0$ admits an eight-dimensional Lie
symmetry algebra
 generated by the following operators\footnote{We remark that in the case
 $A=0$ the linearizing transformation is obtained by
means of the canonical representation of a two-dimensional abelian intransitive
subalgebra \cite{Lie12}. One such subalgebra is that generated by $\Gamma_7$
and $\Gamma_8$ and therefore the transformation
$$ \hat t=\tan(\omega t),\quad \quad \hat x =\frac{h}{\cos(\omega t)}$$
takes equation \eqref{lieniilinom} into the free particle, while transformation
\eqref{lintoho} takes equation \eqref{lieniilinom} into the harmonic oscillator
with frequency $\omega$.}
\begin{eqnarray}
\Gamma_{1} &=&\partial_{t},\nonumber
\\
\Gamma_{2} &=&\cos(2  \omega t) \partial_{t}- \omega  \frac{h}{h'} \sin(2
\omega t) \partial_{x},\nonumber
\\
\Gamma_{3} &=&\sin(2  \omega t) \partial_{t}+ \omega  \frac{h}{h'} \cos(2
\omega t) \partial_{x},\nonumber
\\
\Gamma_{4} &=& \frac{h}{\omega^2} \cos( \omega t) \partial_{t} -
\frac{h^2}{\omega h'} \sin( \omega t) \partial_{x},\nonumber
\\
\Gamma_{5} &=&\frac{h}{\omega^2} \sin( \omega t) \partial_{t} +
\frac{h^2}{\omega h'} \cos( \omega t) \partial_{x},
\\
\Gamma_{6} &=&\frac{h}{h'} \partial_{x},\nonumber
\\
\Gamma_{7} &=&\frac{\omega^2}{h'} \sin( \omega t) \partial_{x},\nonumber
\\
\Gamma_{8} &=&\frac{\omega^2}{h'} \cos( \omega t) \partial_{x},\nonumber
\label{symmlienii}
\end{eqnarray}
while  equation \eqref{lieniilinom} with $A\neq 0$
 admits a three-dimensional Lie symmetry algebra admitted generated
  by $\Gamma_1,\Gamma_2,\Gamma_3$, i.e.:
 \begin{eqnarray}
\Gamma_{1} &=&\partial_{t},\nonumber
\\
\Gamma_{2} &=&\cos(2  \omega t) \partial_{t}- \omega  \frac{h}{h'} \sin(2
\omega t)
\partial_{x},
\\
\Gamma_{3} &=&\sin(2  \omega t) \partial_{t}+ \omega  \frac{h}{h'} \cos(2
\omega t)
\partial_{x}.\nonumber
\label{symmlienii_A}
\end{eqnarray}
 The three symmetries \eqref{symmlienii_A} are a representation of the
complete symmetry group of equation \eqref{lieniilinom}, namely a group that
completely specifies a given differential equation through its algebraic
representation \cite{Krause}. Indeed if we impose to the following general
second-order ordinary differential equation \begin{equation}\ddot{x}(t)=
{F}(t,{x},{\dot{x}}) \end{equation} to admit the symmetry algebra with
generators \eqref{symmlienii_A}, then we obtain equation \eqref{lieniilinom}, a
family of equations characterized by the parameter $A$.\\ We now show that
equation \eqref{lieniilinom} hides linearity. If we solve equation
\eqref{lieniilinom} with respect to $A$ and derive once with respect to $t$,
then we obtain the following third-order equation\footnote{This method is
described in \cite{Leach03}.}
\begin{equation}
\dddot x= - 3\left(\frac{h''}{h} +\frac{h'}{h}\right)\dot x\ddot x
-\left(\frac{h'''}{h'}+ 3\frac{h''}{h}\right)\dot x^3 - 4\omega^2\dot x,
\end{equation}
that admits a seven-dimensional Lie symmetry algebra\footnote{The seven
generators are $\Gamma_1,\Gamma_2,\Gamma_3,\Gamma_6$ in \eqref{symmlienii} and
  $$ X_1=\frac{1}{hh'}
\partial_{x}, \quad X_2=\frac{1}{hh'} \cos( 2\omega t) \partial_{x},
 \quad X_3=\frac{1}{hh'} \sin( 2\omega t) \partial_{x}.$$ }
 and therefore is linearizable \cite{Lie12}. The linearizing
transformation, that is obtained by means of the canonical representation of a
two-dimensional abelian intransitive subalgebra\footnote{Namely that generated
by $X_2$ and $X_3$.} \cite{Lie12}, is
\begin{equation}
\tilde t=\tan(2\omega t),\quad \quad  \tilde x =\frac{h^2}{2\cos(2\omega t)},
\end{equation}
that yields the following linear equation
\begin{equation}
\frac{{\rm d}^3 \tilde x}{{\rm d}\tilde t^3}=-\frac{3\tilde t}{1+\tilde t^2}
\frac{{\rm d}^2 \tilde x}{{\rm d}\tilde t^2}.
\end{equation}
%df(ut,yt,3)=-3*yt*df(ut,y,2)/(1+yt^2);
In particular the transformation $u =\displaystyle\frac{h^2}{2}$ yields
\begin{equation}
\dddot u=-4\omega^2 \dot u, \label{2omdeq}
\end{equation}
namely the once-derived linear harmonic oscillator with frequency $2\omega$.\\

In this paper we apply the quantization algorithm that preserves the Noether
symmetries to equation \eqref{lieniilinom} in the case  $A=0$ and in the case
$A\neq0$, and compare our findings with those in \cite{Partha2013}. We also
determine the eigenvalues and the eigenfunctions of the obtained \SE~using its
Lie symmetries i.e. the method developed in a series of papers
\cite{Leach05,KostisLeach05,jlmschqm,Nucci2010}.

\section{Quantization of the  Li\'enard II equation \eqref{lieniilinom}}
\label{lienardii}\noindent We quantize  equation \eqref{lieniilinom} by
considering first the case $A=0$, i.e.:
\begin{equation}
\ddot{x} + \frac{h''}{h'}  \dot{x}^2 + \omega^2 \frac{h}{h'}=0
\label{lieniilinom_l}
\end{equation}
 and then the case $A\neq 0$.

\subsection{Equation \eqref{lieniilinom_l}}
As shown in \cite{LakshmananLienII2013} equation \eqref{lieniilinom_l} is
linearizable, therefore in order to quantize it we follow the three Steps
\cite{GN_liensch} as recalled in the Introduction.\\

\noindent {\bf Step 1.} The transformation that takes equation
\eqref{lieniilinom_l} into the harmonic oscillator with frequency $\omega$ is
\eqref{lintoho}
\cite{LakshmananLienII2013}. \\

\noindent {\bf Step 2.} The Lagrangian \eqref{lagrlienii} corresponding to
equation \eqref{lieniilinom_l} is
\begin{equation}
L_0 = \half(h')^{2} \dot{x}^2 -\half \omega^2 h^2. \label{lagrlieniilinom}
\end{equation}
It admits five Noether point symmetries, namely $\Gamma_1$, $\Gamma_2$,
$\Gamma_3$, $\Gamma_7$ and $\Gamma_8$ in \eqref{symmlienii}.

\begin{remark}
The Jacobi Last Multiplier \eqref{jlmej} in the case of equation
\eqref{lieniilinom_l} becomes \begin{equation} M = (h')^{2}.\label{jlmlin}
\end{equation} Since there is a link between Jacobi Last Multiplier and Lie symmetries
 \cite{Lie1874,jlm05} then it is interesting to underline that
the Jacobi Last Multipliers \eqref{jlmlin} can be obtained by means of the two
symmetries $\Gamma_{7}$ and $\Gamma_{8}$, that generate  an intransitive
two-dimensional abelian subalgebra yielding the linear transformation
\eqref{lintoho}. In fact  the Jacobi Last Multiplier \eqref{jlmlin} is the
reciprocal of the determinant:
\begin{equation}
\Delta_{78} = {\rm det} \left [\begin {array} {ccc}
1 & \dot x & -\displaystyle\frac{h''}{h'}  \dot{x}^2 - \omega^2 \displaystyle\frac{h}{h'}\\
0 & \displaystyle\frac{\omega^2}{h'} \sin( \omega t)
& \displaystyle\omega^2 \frac{\cos(\omega  t) \omega h'-\sin(\omega  t) \dot{x} h''}{(h')^2}\\
0 & \displaystyle\frac{\omega^2}{h'} \cos( \omega t) & \displaystyle-\omega^2
\frac{\sin(\omega  t) \omega h'-\cos(\omega  t) \dot{x} h''}{(h')^2}
\end{array} \right]
= - \frac{\omega^{5}}{(h')^2},
\end{equation}
apart from the unessential constant factor $-\omega^5$.
\end{remark}

\noindent {\bf Step 3.} The Schr\"{o}dinger equation of the linear harmonic
oscillator with frequency $\omega$ is:
\begin{equation}
2 \imath \psi_{t} + \psi_{\xi\xi} - \omega^2 \xi^2\psi=0
\end{equation}
with $\psi=\psi(t,\xi)$. If we  apply the transformation \eqref{lintoho}, then
we obtain the  \SE~   of equation \eqref{lieniilinom_l}:
\begin{equation}
2 \imath \psi_{t} + \frac{\psi_{xx}}{(h')^2} -\frac{h''\psi_{x}}{(h')^3}
-\omega^2 h^2 \psi=0. \label{lieniischr}
\end{equation}
We now check the classical consistency of the Schr\"{o}dinger equation
\eqref{lieniischr}.  Using the REDUCE programs \cite{Nucci1996} we find that
its Lie point symmetries are generated by the following operators:
\begin{eqnarray}
\Xi_{1} &=& \Gamma_1,\nonumber\\
\Xi_{2} &=& \Gamma_{2} +\frac{\omega \psi}{2} \left(\sin(2 \omega t)
- 2 \cos(2 \omega t) h^2 \imath \omega\right) \partial_{\psi},\nonumber\\
\Xi_{3} &=& \Gamma_{3} -\frac{\omega \psi}{2} \left(\cos(2 \omega t) + 2 \sin(2
\omega t) h^2 \imath \omega\right) \partial_{\psi},
\\
\Xi_{4} &=& \Gamma_{7}
+\cos(\omega t) h \imath \omega \psi \partial_{\psi},\nonumber\\
\Xi_{5} &=& \Gamma_{8} - \sin(\omega t) h \imath \omega \psi
\partial_{\psi}.\nonumber
 \label{schrsymmetries}
\end{eqnarray}
and the two homogeneity and linearity symmetries.

In \cite{Leach05,KostisLeach05,jlmschqm,Nucci2010} and more recently in
\cite{Nucci2013,GN_liensch} it was shown how to find the eigenfunctions and the
eigenvalues of the Schr\"{o}dinger equation by means of its admitted Lie
symmetries.

We apply this method to the Schr\"{o}dinger equation \eqref{lieniischr}.

Let us rewrite the Lie point symmetries \eqref{schrsymmetries} of equation
\eqref{lieniischr} in complex form, i.e.:
\begin{eqnarray}
\Omega_{1} &=& \imath \partial_{t},\nonumber\\
\Omega_{2\pm} &=& e^{\pm 2\imath \omega t} \left[
\partial_{t} \pm \imath \frac{h}{h'}\partial_{x}
-\imath \left(\omega^2 h^2  \pm \frac{\imath}{2}\omega \right)\psi
\partial_{\psi} \right],
\\
\Omega_{3\pm} &=& e^{\pm\imath \omega t}
\left(\frac{1}{h'}\partial_{x}\mp\omega h \psi\partial_{\psi}\right). \nonumber
\label{schsymI}
\end{eqnarray}

The operators $\Omega_{3\pm}$ act as creation and annihilation operators. In
fact if we consider the invariant surfaces associated with these two operators:
\begin{equation}
\Omega_{3\pm} F(t,x,\psi) = 0,
\end{equation}
one gets
\begin{equation}
F(t,x,\psi) = f(t,\psi e^{\pm\frac{\omega}{2}h^2})
\end{equation}
namely  the following similarity solutions of \eqref{lieniischr}:
\begin{equation}
\psi_{\pm} = T_{\pm}(t) e^{\mp\frac{\omega}{2}h^2}.
\end{equation}
Since we want a solution that goes to zero at infinity, then we have to choose
the invariant surface relative to $\Omega_{3+}$. Hence the operator
$\Omega_{3-}$ acts  as a creation operator, while $\Omega_{3+}$ as an
annihilation one. Substituting $\psi_{+}$ into the \SE~\eqref{lieniischr}
yields $T_{+}=e^{-\frac{\imath}{2} \omega t}$, thus the ground
state\footnote{Apart from an unessential normalization constant.}
 is:
\begin{equation}
\psi_{0} = e^{-\frac{\imath}{2} \omega t -\frac{\omega}{2}h^2}. \label{psi0}
\end{equation}
The operator $\Omega_{1}$ acts as an eigenvalue operator:
\begin{equation}
\Omega_{1} \psi_{0} = \frac{\omega}{2} \psi_{0},
\end{equation}
since it yields the ground state energy $E_{0}=\omega/2$, that corresponds to
the quantum harmonic oscillator.

We use the creation operator $\Omega_{3-}$ and the linearity
operator\footnote{Namely $ \Omega_{\chi}=\chi\partial_{\psi}$ with $\chi$ any
solution of \eqref{lieniischr}.} $\Omega_{\psi_{0}}$ in order to construct the
higher states. Since the commutator:
\begin{equation}
\left[\Omega_{3-},\Omega_{\psi_{0}}\right] = -2\omega h e^{-\frac{3\imath}{2}
\omega t -\frac{\omega}{2}h^2}.
\end{equation}
then:
\begin{equation}
\psi_{1} = -2\omega h e^{-\frac{3\imath}{2} \omega t -\frac{\omega}{2}h^2}.
\end{equation}
is another solution of \eqref{lieniischr} which satisfies the proper boundary
conditions and have eigenvalue:
\begin{equation}
\Omega_{1} \psi_{1} = \frac{3\omega}{2} \psi_{0},
\end{equation}
i.e. $E_{1}=3\omega/2>E_{0}$.

If we evaluate the commutator:
\begin{equation}
\left[\Omega_{3+},\Omega_{\psi_{1}}\right] = -2\omega \Omega_{\psi_{0}},
\end{equation}
we have a multiple of the ground state \eqref{psi0}.

Iterating this process yields all the eigenfunctions, i.e.:
\begin{equation}
\left[\Omega_{3-},\Omega_{\psi_{n-1}}\right] = \psi_{n} \partial_{\psi} =
\Omega_{\psi_{n}}. \label{phin}
\end{equation}
The generic eigenvalue and eigenfunction can be derived in the following
manner. We evaluate the commutator between $\Omega_{3-}$ and $\Omega_{\chi}$,
where $\chi$ is a generic solution of \eqref{lieniischr} i.e.
\begin{equation}
\left[\Omega_{3-},\Omega_{\chi}\right] = e^{-\imath \omega t}
\left(\frac{1}{h'}\partial_{x} -\omega h\right)\chi\partial_{\psi},
\end{equation}
and then we generate the $n$th eigenfunction by using the iteration procedure
\eqref{phin}, i.e.:
\begin{equation}
\begin{aligned}
\psi_{n} &= e^{-\imath \omega t} \left(\frac{1}{h'}\partial_{x} -\omega
h\right) \psi_{n-1}
\\
&=e^{-2\imath \omega t} \left(\frac{1}{h'}\partial_{x} -\omega h\right)^2
\psi_{n-2}
\\
&=\vdots
\\
&=e^{-n \imath \omega t} \left(\frac{1}{h'}\partial_{x} -\omega
h\right)^{n}\psi_{0}
\\
&=e^{-\left(n+\half\right)\imath\omega t} \left(\frac{1}{h'}\partial_{x}
-\omega h\right)^{n} \left(e^{-\frac{\omega}{2}h^2}\right)
\\
&=(-1)^{n}\omega^{\frac{n}{2}}e^{-\left(n+\half\right)\imath\omega t}
H_{n}(\sqrt{\omega}h)e^{-\frac{\omega}{2}h^2}.
\end{aligned}
\end{equation}
where $H_{n}$ is the $n$-th Hermite polynomial \cite{Abramowitz1964}. This
trivially gives the $n$-th eigenvalue:
\begin{equation}
\Omega_{1} \psi_{n} = \omega \left(n +\half\right) \psi_{n}.
\end{equation}
Thus we have proven that the spectrum of the Schr\"{o}dinger equation
\eqref{lieniischr} is equal to that of the quantum harmonic oscillator of
frequency $\omega$.

\subsection{Equation \eqref{lieniilinom}}
Equation \eqref{lieniilinom} is not linearizable by means of a point
transformation. In fact it admits just three symmetries \eqref{symmlienii_A},
that generates an algebra $sl(2,\R)$, therefore we cannot use the three Steps
1,2,3 as in the case $A=0$. Instead we have to use the more general method
proposed in \cite{gallipoli10} and recalled in the Introduction, namely Steps
I,II,III, that is to find a Lagrangian admitting those three symmetries as
Noether symmetries, and then use them to quantize equation
\eqref{lieniilinom}.\\

\noindent {\bf Step I.} The three-dimensional Lie symmetry algebra of equation
\eqref{lieniilinom} is generated by the operators \eqref{symmlienii_A}.\\

\noindent {\bf Step II.} The Lagrangian \eqref{lagrlienii} corresponding to
equation \eqref{lieniilinom} is:
\begin{equation}
L_A=\frac{(h')^2}{2}\dot x^2 + \frac{A}{2h^2} - \frac{\omega^2}{2}h^2.
\label{lagrlieniilinom_A}
\end{equation}
It admits the three symmetries \eqref{symmlienii_A} as Noether symmetries.\\

\noindent {\bf Step III.} We now construct the Schr\"odinger equation admitting
the Noether symmetries \eqref{symmlienii_A} as Lie symmetries. We begin with
equation \eqref{sch}, i.e.:
\begin{equation} 2i\psi_t+ f_{11}(x)\psi_{xx}+h_1(x)\psi_x
+f_0(x)\psi=0,\end{equation} and thus we obtain that the Schr\"odinger equation
corresponding to the equation \eqref{lieniilinom} is\footnote{Introducing
$\hbar$ into \eqref{schlienii_A}, i.e.
$$  2i\hbar\psi_t+ \hbar^2\frac{\psi_{xx}}{(h')^2}-\hbar^2\frac{h''\psi_{x}}{(h')^3}
+\left(\frac{A}{h^2}-\omega^2h^2\right)\psi=0,$$ and performing the classical
limit \cite{Landau} one indeed obtains equation \eqref{lieniilinom}.}
\begin{equation} 2i\psi_t+ \frac{\psi_{xx}}{(h')^2}-\frac{h''\psi_{x}}{(h')^3}
+\left(\frac{A}{h^2}-\omega^2h^2\right)\psi=0,
\label{schlienii_A}\end{equation} and its Lie symmetries are generated by the
operators $\Xi_1,\Xi_2,\Xi_3$ in \eqref{schrsymmetries} and the usual
homogeneity and linearity symmetries.\\

As before, the eigenfunctions and the eigenvalues of the Schr\"{o}dinger
equation \eqref{schlienii_A} can be derived by means of its admitted Lie
symmetries. However here the operators $\Omega_{2\pm}$ in \eqref{schsymI} act
as creation and annihilation operators. Since the procedure is the same we do
not write down all the details. We have obtained that the eigenfunctions are:
\begin{equation}
\psi_n= h^{\frac{k+1}{2}}e^{-\imath\frac{k+2}{2} \omega t
-\frac{\omega}{2}h^2}L_n^{k/2}\left(\omega h^2\right), \label{psin_A}
\end{equation}
with $k=\sqrt{1-4A}$, and $L_n^{k/2}$ the associated Laguerre polynomials,
while the energy eigenvalues are:
\begin{equation}
E_n=2 \omega\left(n+\frac{1}{2}+ \frac{k}{4}\right).\label{En}
\end{equation}
It is not a surprise that the energy eigenvalues \eqref{En} are exactly those
of the quantum isotonic oscillator \cite{Goldman}.
 We also underline that they correspond to
  those of the quantum harmonic oscillator with frequency $2\omega$.

\section{Comparison with the quantization  in \cite{Partha2013}}

In \cite{Partha2013} the von Roos' ordering method
\cite{vonRoos1983,vonRoos1985} for position dependent masses was applied to
\eqref{lienii} and the following \SE~ was obtained\footnote{In
\cite{Partha2013} the time-independent Schr\"{o}dinger equation was derived.} :
\begin{equation}
2\imath \phi_{t} +e^{2\int^{x}f(\xi)\ud \xi} \left\{\phi_{xx}- 2 f\phi_{x} +
\left[(\beta+1)(2 f^2 -f')+ 4\alpha(\alpha+\beta+1)f^2\right]\phi \right\} -
V\phi, \label{schrpartha}
\end{equation}
with
\begin{equation}
V=\int^{x} e^{2 \int^{\xi}f(\eta)\ud \eta}g(\xi)\ud \xi,
\end{equation}
and such that the parameters $\alpha$ and $\beta$ are related to another
parameter $\gamma$ such that $\alpha + \beta + \gamma=-1$.

Then the authors factorized the wave function $\phi$ \cite{Bhattacharjie1962},
i.e.:
\begin{equation}
\phi(t,x)= e^{-\imath E t} w(x) G(u(x)),
\end{equation}
with the function $G$ being the solution of a generic second-order linear
differential equation:
\begin{equation}
\frac{\ud^{2} G}{\ud u^2} + Q(u) \frac{\ud G}{\ud u} + R(u)G(u)=0.
\label{sturm}
\end{equation}
In \cite{Partha2013}, different differential equations
 were taken into consideration, in particular  the
 Hermite and the associated Laguerre equation.\\

If equation \eqref{sturm} is the Hermite differential equation, then in
\cite{Partha2013} it was derived that  $\alpha=\gamma-1/4$, $\beta=-1/2$, and
the potential $V$ had to be
\begin{equation}
V=\half\left(\int^{x}e^{\int^{\xi}f(\eta)\ud \eta}\ud\xi\right)^2.
\label{hermvg}
\end{equation}
We substitute  $f=h''/h'$ into equation \eqref{schrpartha} and find to yield
the \SE~ \eqref{lieniischr} that we have derived with $\omega=1$ and
$\phi=\sqrt{h'}\psi$.

We underline that the \SE~ \eqref{schrpartha} with the potential $V$ given in
\eqref{hermvg} admits the symmetries $\Xi_{i}$, $i=1,\ldots,5$ in
\eqref{schrsymmetries} if and only if $\alpha=\gamma-1/4$, $\beta=-1/2$.\\

If equation \eqref{sturm} is the associated Laguerre equation, then in
\cite{Partha2013} it was derived that $\alpha=\gamma-1/4$, $\beta=-1/2$, and
the potential $V$ had to be:
\begin{equation}
V=\frac{1}{8}\left(\int^{x}e^{\int^{\xi}f(\eta)\ud \eta}\ud\xi\right)^2
+ \frac{\displaystyle \half\left[\frac{3}{4}-(1+\mu)(1-\mu)\right]}%
{\displaystyle\left(\int^{x}e^{\int^{\xi}f(\eta)\ud \eta}\ud\xi\right)^2}.
\label{laguerrev}
\end{equation}
We substitute  $f=h''/h'$ into equation \eqref{schrpartha} and find to yield
the \SE~ \eqref{schlienii_A} that we have derived with $\omega=\frac{1}{2}$ and
$\phi=\sqrt{h'}\psi$.

\section{Final remarks}
A new algorithm for quantization  that requires the preservation of Noether
symmetries in the passage from classical to quantum mechanics\footnote{Namely,
 the derived Schr\"{o}dinger equation is such that the
independent-variables part of its admitted Lie symmetries corresponds to the
Noether symmetries of the classical equations.} has been recently introduced
and applied to both one-dimensional and two-dimensional Lagrangian equations
\cite{gallipoli10,Bregenz11,Nucci2012,AGMP09,GN_liensch,PMNP13}.

In this paper we have  applied this new method to the quadratic Li\'enard-type
equation \eqref{lieniilinom}, both in the linearizable an non-linearizable
case, and compared our results with that determined in \cite{Partha2013}. We
have found that the Schr\"{o}dinger equations obtained in \cite{Partha2013} can
be determined by means of the quantization that preserves the Noether
symmetries.

Even in quantum mechanics whenever differential equations are involved, Lie and
Noether symmetries have a fundamental role: Noether symmetries yield the
correct Schr\"odinger equation and its Lie symmetries can be algorithmically
used to find the eigenvalues and eigenfunctions.

\section*{Acknowledgement}
MCN acknowledges the support of the Italian Ministry of University and
Scientific Research through PRIN 2010-2011, Prot. 2010JJ4KPA\_004.


\begin{thebibliography}{10}
\bibitem{Abramowitz1964}
M.~Abramowitz and I.~A.~Stegun.
\newblock {\em Handbook of mathematical functions with formulas, graphs, and
  mathematical tables}, volume~55 of {\em National Bureau of Standards Applied
  Mathematics Series}.
\newblock U.S. Government Printing Office, Washington, D.C., 1964.

\bibitem{Bhattacharjie1962}
A.~Bhattacharjie and E.~C.~G.~Sudarshan.
\newblock A class of solvable potentials.
\newblock {\em Nuovo Cimento}, 25(4):864--879, 1962.
\bibitem{Bjorken1964}
D. Bjorken and S.~D. Drell.
\newblock {\em Relativistic Quantum Mechanics}.
\newblock McGraw-Hill Book Co., New York, 1964.

\bibitem{Chandrasekar2012}
V.~K.~Chandrasekar, J.~H. Sheeba, R.~G.~Pradeep, R.~S.~Divyasree, and
  M.~Lakshmanan.
\newblock A class of solvable coupled nonlinear oscillators with amplitude
  independent frequencies.
\newblock {\em Physics Letters A}, 376(32):2188--2194, 2012.

\bibitem{Partha2013}
A.~G.~Choudhury and P.~Guha.
\newblock Quantization of the Li\'enard ii equation and Jacobi's last
  multiplier.
\newblock {\em J. Phys. A: Math. Theor.}, 46(16):165202, 2013.

\bibitem{Euler1769}
L.~Euler.
\newblock {\em (E366) Institutionum calculi integralis, Volumen Secundum}.
\newblock Petropoli impensis academi\ae imperialis scientiarum, Petersburg,
  1769.

\bibitem{Goldman}
I.~I.~ Goldman and V.~D.~ Krivchenkov. {\em Problems in Quantum Mechanics}.
Pergamon Press, London, 1961.

\bibitem{Goldstein80}
H.~Goldstein.
\newblock {\em Classical {Mechanics}}.
\newblock Addison-Wesley, Reading (MA), 2nd edition, 1980.


\bibitem{Gonzalez1983}
F.~Gonz\'{a}lez-Gasc\'{o}n and A.~Gonz\'{a}lez-L\'{o}pez.
\newblock Symmetries of differential equations.
\newblock {\em J. Math. Phys.}, 23:2006--2021, 1983.

\bibitem{Gonzalez1988}
Gonz\'{a}lez-L\'{o}pez A.
\newblock Symmetries of linear systems of second-order ordinary differential
  equations.
\newblock {\em J. Math. Phys.}, 29:1097--1105, 1988.


\bibitem{GN_liensch}
G.~Gubbiotti and M.~C. Nucci.
\newblock Noether symmetries and the quantization of a {Li\'enard}-type
  nonlinear oscillator.
\newblock {\em J. Nonlinear Math. Phys.}, 21:248--264, 2014.


\bibitem{Jacobi45}
C.~G.~J. Jacobi.
\newblock Theoria novi multiplicatoris systemati {\ae}quationum differentialium
  vulgarium applicandi.
\newblock {\em J. Reine Angew. Math.}, 29:213--279 and 333--376, 1845.

\bibitem{Krause}
J.~Krause, On the complete symmetry group of the classical Kepler system. {\em
J. Math. Phys.}, 35:5734--5748, 1994.

\bibitem{Landau}
L.~D.~Landau and E.~M.~Lifshitz. {\em Quantum mechanics. Non-relativistic
theory Vol.3.} Pergamon Press, Oxford, 1991.


\bibitem{Leach03}
P.~G.~L. Leach. Equivalence classes of second-order ordinary differential
equations with only a three-dimensional Lie algebra of point symmetries and
linearisation. {\em J. Math. Anal. Appl.}, 284:31-–48, 2003.
\bibitem{Leach05}
P.~G.~L. Leach.
\newblock The solution of some quantum nonlinear oscillators with the common
  symmetry group {$\mathrm{SL}(2,\R)$}.
\newblock {\em J. Phys. A: Math. Gen.}, 38:1543--1552, 2005.
\bibitem{KostisLeach05}
P.~G.~L. Leach and K.~Andriopoulos.
\newblock Wave-functions for the time-dependent linear oscillator and Lie point
  symmetries.
\newblock {\em J. Phys. A: Math. Gen.}, 38:4365--4374, 2005.

\bibitem{Lie1874}
S.~Lie.
\newblock Veralgemeinerung und neue {V}erwerthung der {J}acobischen
  {M}ultiplicator-{T}heorie.
\newblock {\em Forhandlinger I Videnskabs-Selskabet I Christiania}, pp.
  255--274, 1874.

\bibitem{Lie12}
S.~Lie.
\newblock {\em Vorlesungen \"uber {D}ifferentialgleichungen mit bekannten
  infinitesimalen {T}ransformationen}.
\newblock Teubner, Leipzig, 1912.

\bibitem{Louisell1990}
W.~H. Louisell.
\newblock {\em Quantum Statistical Properties of Radiation}.
\newblock John Wiley \& Sons, New York, 1990.

\bibitem{Nucci1996}
M.~C. Nucci. \newblock Interactive REDUCE programs for calculating Lie point,
non-classical, Lie-B\"{a}cklund, and approximate symmetries of differential
equations: manual and floppy disk.\newblock In {\em CRC Handbook of Lie Group
Analysis of Differential Equations. Vol. 3: New Trends in Theoretical
Developments and Computational Methods}, N.H. Ibragimov ed., CRC Press, Boca
Raton (1996) pp. 415--481.

\bibitem{jlm05}
M.C. Nucci. Jacobi last multiplier and Lie symmetries:
 a novel application of an old relationship.
  {\em J. Nonlinear Math. Phys.}, 12:284--304, 2005.

\bibitem{gallipoli10}
M.~C. Nucci.
\newblock Quantization of classical mechanics: shall we {L}ie?
\newblock {\em Theor. Math. Phys.}, 168:997--1004, 2011.

\bibitem{Bregenz11}
M.~C. Nucci.
\newblock From Lagrangian to Quantum Mechanics with Symmetries.
\newblock {\em J. Phys.: Conf. Ser.}, 380:012008, 2012.

\bibitem{Nucci2012}
M.~C. Nucci.
\newblock Quantizing preserving {N}oether symmetries.
\newblock {\em J. Nonlinear Math.
Phys.}, 20:451--463, 2013.
\bibitem{AGMP09}
M.~C. Nucci. Symmetries for thought. {\em Math. Notes Miskolc}, 14:461--474,
2013.
\bibitem{PMNP13}
M.~C. Nucci. Spectral realization of the Riemann zeros by quantizing $H =
w(x)(p + \ell_p^2/p)$: the Lie-Noether symmetry approach. {\em J. Phys.: Conf.
Ser.}, 482:012032, 2014.

\bibitem{jlmschqm}
M.~C. Nucci and P.~G.~L. Leach.
\newblock Gauge variant symmetries for the {Schr\"odinger} equation.
\newblock {\em Nuovo Cimento B}, 123:93--101, 2008.

\bibitem{Nucci2010}
M.~C. Nucci and P.~G.~L. Leach.
\newblock An algebraic approach to laying a ghost to rest.
\newblock {\em Phys. Scripta}, 81(5):055003, 2010.

\bibitem{laggal}
M.C. Nucci and P.G.L. Leach, Lagrangians galore, {\em J. Math. Phys.} {\bf 48},
123510 (2007)

\bibitem{CP07Rao1JMP}
 M.C. Nucci and P.G.L. Leach, Jacobi last multiplier and Lagrangians
for multidimensional linear systems,  {\em J. Math. Phys. 49} (2008)  073517

\bibitem{Nucci2013}
M.C. Nucci and P.G.L. Leach.
\newblock Lie groups and quantum mechanics.
\newblock {\em J. Math. Anal. Appl.}, 406:219--228, 2013.

\bibitem{c-iso}
M.~C. Nucci, P.~G.~L. Leach, and K.~Andriopoulos.
\newblock Lie symmetries, quantisation and $c$-isochronous nonlinear
  oscillators.
\newblock {\em J. Math. Anal. Appl.}, 319:357--368, 2006.


\bibitem{nuctam_1lag}
M.C. Nucci and K.M. Tamizhmani.
\newblock Using an old method of {J}acobi to derive {L}agrangians: a nonlinear
  dynamical system with variable coefficients.
\newblock {\em Nuovo Cimento B}, 125:255--269, 2010.

\bibitem{LakshmananLienII2013}
A.~K.~Tiwari, S.~N.~Pandey, M.~Senthilvelan, and M.~Lakshmanan.
\newblock Classification of Lie point symmetries for quadratic {Li\'enard} type
  equation $\ddot{x}+f(x)\dot{x}^2+g(x)=0$.
\newblock {\em J. Math. Phys.}, 54:053506, 2013.

\bibitem{vonRoos1983}
O.~von Roos.
\newblock Position-dependent effective masses in semiconductor theory.
\newblock {\em Phys. Rev. B}, 27:7547--7552, Jun 1983.

\bibitem{vonRoos1985}
O.~von Roos and H.~Mavromatis.
\newblock Position-dependent effective masses in semiconductor theory. {II}.
\newblock {\em Phys. Rev. B}, 31:2294--2298, Feb 1985.


\bibitem{Weyl1927}
H.~Weyl.
\newblock Quantenmechanik und gruppentheorie.
\newblock {\em Zeitschrift f\"{u}r Physik}, 46:1--46, 1927.

\bibitem{Whittaker}
E.T. Whittaker, {\it A Treatise on the Analytical Dynamics of Particles and
Rigid Bodies}, Cambridge University Press, Cambridge  (1988)
\end{thebibliography}
\end{document}